\documentstyle[preprint,prl,aps]{revtex}    
\begin{document}

\newcommand \be  {\begin{equation}}
\newcommand \bea {\begin{eqnarray} \nonumber }
\newcommand \ee  {\end{equation}}
\newcommand \eea {\end{eqnarray}}

 \title{\bf COMMENT ON ``TURBULENT CASCADES IN FOREIGN EXCHANGE MARKETS''}

\vskip 3 true cm

\author{Alain Arneodo$^1$, Jean-Philippe Bouchaud$^{2,3}$, Rama Cont$^{3,5}$, 
Jean-Fran\c{c}ois Muzy$^1$,\\
Marc Potters$^3$ and Didier Sornette$^{3-5}$}

\date{\it
$^1$ Centre de Recherche Paul Pascal, Av. Schweitzer, 33600 Pessac,
France\\
$^2$ Service de Physique de l'\'Etat Condens\'e,
 Centre d'\'etudes de Saclay, \\ Orme des Merisiers, 
91191 Gif-sur-Yvette Cedex, France \\
$^3$ Science \& Finance, 109--111, rue Victor Hugo, 92535 Levallois Cedex,
France\\
$^4$ Department of Earth and Space Science and Institute of Geophysics
and Planetary Physics\\ University of California, Los Angeles,
California 90095\\
$^5$ Laboratoire de
Physique de la Mati\`ere Condens\'ee, CNRS URA190\\ Universit\'e des
Sciences, B.P. 70, Parc Valrose, 06108 Nice Cedex 2, France
}

\maketitle

\vskip 3 true cm

Recently, Ghashghaie et al.\ have shown that some statistical
aspects of fully developed turbulence and exchange rate fluctuations
exhibit striking similarities \cite{Peinke}. The authors then suggested
that the two problems might be deeply connected, and speculated on the
existence of an `information cascade' which would play the role in
finance of the well known Kolmogorov energy cascade in turbulence
\cite{Frisch}.

Here we want to convince the reader that the two problems differ on a
fundamental aspect, namely, correlations. Spatial correlations lead to
the famous $-\frac{5}{3}$ power-law for the spectrum of the velocity
fluctuations \cite{Frisch}, cut-off on the low frequency side by the
energy injection mechanism, and on the high frequency side by
dissipation \cite{Peinke}. No temporal correlations of the sort are
visible in the power-spectrum of financial time series --- see Fig.\
1, corresponding to the same data as studied in \cite{Peinke}, where a
slope of $-2$ is observed on a log-log scale. As seen in the inset,
this corresponds to a totally `white' spectrum for the price change
signal (no correlations).  If such correlations existed, by the way,
it would be rather easy to use them to earn money!

The quality of the `log-normal' fit reported in \cite{Peinke} can only
be suggestive, since it is a two parameter fit {\it for each time
delay} $\Delta t$. This must be compared to another proposal, which is
that the fluctuations of financial assets are well described by a {\it
truncated L\'evy process}\cite{Koponen}, or some other fat-tail
process, with {\it independent}\/ increments. This was originally
suggested in \cite{Stanley} where the S\&P 500 index was studied, and
later substantiated on many other financial time series, using
independent techniques (wavelets \cite{Arneodo}, or direct
histogramming and convolution [M. Potters, unpublished data]). In
Fig.\ 2, we show a `truncated L\'evy' fit of the data. There are three
parameters, but which are once and for all fixed on the smallest time
scale $\Delta t=5$ min, while the larger time scales are obtained by
convolution which amounts to assume independence of the
increments. Notice that this method predicts both the shape and the
overall scale of the distribution.

The mechanism by which fat tails in financial data disappear as the
time scale grows is nothing but the consequence of the central limit
theorem (see e.g.\ \cite{GK,Newton}), which can safely be applied when
these tails decay sufficiently fast and when correlations are absent.
Subtle temporal correlations are actually observed in the evolution of
the `volatility', i.e.\ the amplitude of the fluctuations, leading to
deviations from a simple convolution rule for the distribution of
increments. However these deviations remain small and still allow for
the application of the central limit theorem.  On the other hand, the
core of the problem in turbulence is precisely the existence of very
strong correlations preventing the use of the central limit theorem!

\noindent
{\bf Acknowledgments.}
We are very grateful to Y. Gagne for the permission to use
his experimental turbulent signals.

\begin{figure}
\caption{Fourier transform of the price autocorrelation function
$\langle x(t) x(t+\tau)\rangle$ as a function of temporal frequency for the
DEM-USD exchange rate (Oct '91--Nov '94) ({\em bottom curve}\/) as
compared to the $-\frac{5}{3}$ power-law spectrum observed for the
spatial velocity fluctuations in turbulent flows (the data were recorded by
Y. Gagne in a wind tunel experiment at $R_{\lambda}= 3050$) ({\em top
curve}\/). Inset: Fourier transform of the price change
autocorrelation function $\langle \Delta x(t) \Delta
x(t+\tau)\rangle$, which is completely flat (and would behave as the
$\frac{1}{3}^{\mbox{\scriptsize th}}$ power of frequency in turbulent flows). Units are arbitrary.}
\end{figure}
\begin{figure}
\caption{Cumulative distribution function of price changes for
different time delays $\Delta t=$5 min, 15 min, 1 hour, 1 day and  5 days
(from bottom to top). Symbols represent empirical data (DEM-USD, Oct
'91--Nov '94): squares for positive price changes ($1-F(\Delta x)$) and
circles for negative ones ($F(-\Delta x)$). The solid line is
a truncated symmetric L\'evy distribution fitted to the data at 5 min
and self-convoluted 3, 12, 84 and 420 times respectively. Note that the
data used correspond to the 7 hour business day in New York.}
\end{figure}

\end{document}